\documentclass[aps,physrev, twocolumn,superscriptaddress]{revtex4-2}
\usepackage{amsmath}

\usepackage[dvipsnames]{xcolor}
\usepackage{lipsum}
\usepackage{amsmath}
\usepackage[colorlinks=true, allcolors=blue]{hyperref}
\usepackage{graphicx}
\usepackage[english]{babel}
\usepackage{dcolumn}
\usepackage{bm}
\usepackage{amssymb}
\usepackage{float}
\usepackage[T1]{fontenc}
\usepackage{physics}
\usepackage{comment}

\begin{document}

\preprint{APS/123-QED}

\title{Spontaneous Vortex Instability in Active Nematics}

\author{Niels de Graaf Sousa}
\author{Kristian Thijssen}
\author{Amin Doostmohammadi}
\email[]{doostmohammadi@nbi.ku.dk}
\affiliation{Niels Bohr Institute, University of Copenhagen, 2100 Blegdamsvej 17, Copenhagen, Denmark}

\date{\today}

\begin{abstract}
     One of the defining results in the study of active matter is the spontaneous flow instability, through which a homogeneous, uniformly aligned state breaks translational symmetry along a single direction and develops sustained flow. The vortex state that emerges at higher activity has instead been attributed to nonlinear dynamics. Using a Floquet-type linear stability analysis, we show that no such mechanism is required: the flowing state undergoes a secondary, zigzag instability that breaks the remaining translational symmetry and produces the vortex state.  We further identify a regime in which the flowing state ceases to exist and vortices emerge directly from the uniformly aligned state. Under channel confinement, the instability selects a length scale that differs from the established active length scale, and sets the number of vortices that appear, leading to a confinement-selected pattern reminiscent of a vortex lattice, opening a route toward direct experimental tests of this instability. Full nonlinear simulations reproduce the predicted onset activities and the selected vortex number.
     
\end{abstract}

\maketitle

Active matter is widely recognized for its ability to generate large-scale patterns from the injection of energy at the individual-particle level. Such patterns have been observed across a range of biological systems, including bacterial suspensions \cite{Goldstein_2004_Bact,bacteria_energy,dunkel2013fluid,bact_NatPhys}, mixtures of cytoskeletal components \cite{MT_Dogic_2012,PRX_turbulence,taming_active_turbulence,MT_toroid,doostmohammadi2018active}, and cell monolayers \cite{cancer_firoblast,AN_Cell_tissues,extrusion_amin_2018}. Over two decades ago, it was hypothesized that active matter could collectively self-organize into a flowing state, driven by the active stress generated by individual constituent particles, with this instability giving rise to spontaneous flows \cite{Simha2002,R_Voituriez_2005}. 

The onset of the flowing instability constitutes one of the characteristic hallmarks of active matter, marking the transition from a quiescent, homogeneous state to one exhibiting spontaneous, self-sustained flow. This instability has served as a paradigmatic example in the theoretical study of active nematics, as it predicts how individually produced energy can create collective patterns and has recently been experimentally observed \cite{Spontanous_transition_exp_2018,flowing_MT}. Many experimental systems are not confined to this one-dimensional geometry, and when the second spatial dimension is unconstrained, the flowing state is generically unstable to further symmetry-breaking perturbations, giving rise to the spontaneous emergence of vortices \cite{Jul_Ramaswamy_2016,living_liquid_crystals} and, at higher activity, more complex spatiotemporal patterns such as active turbulence \cite{Cell_turbulence,dunkel2013fluid,alert2022active,onset_meso_turbulence}. Whereas vortex formation in active nematics has generally been attributed to nonlinear effects, here we show, using a Floquet type perturbation \cite{Cross2009,RevModPhys_Cross}, that this emergence is instead governed by a linear mechanism, arising when the flowing state becomes unstable to perpendicular perturbations that give rise to the zigzag instability. A combined analytical and numerical approach shows that, in an unbounded two-dimensional system, this zigzag instability arises by selecting the largest length scale permitted in the system, in a manner analogous to the flowing instability itself. When the system is confined within a channel, activity selects an intrinsic length scale along the channel direction that differs from the active length scale, thereby setting the number of vortices such that the vortex density scales with activity, and establishing the precursor to the vortex lattice-like pattern \cite{Ceilidh-Dance_2018,MT_exp_channel}.\\

 \begin{figure*}[!t]
    \centering
    \includegraphics[width=\linewidth]{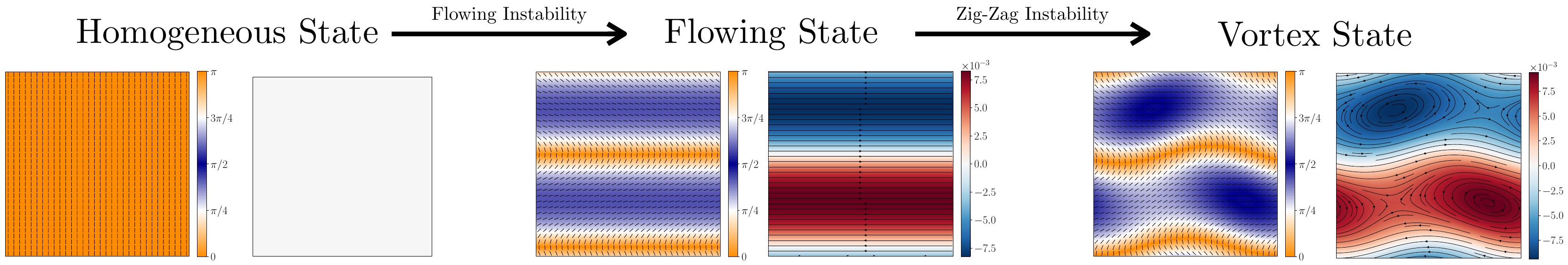}
    \caption{\textbf{Zigzag instability as a precursor to vortex formation in active nematics.} Characteristic director angle (left) and stream function (right) pairs for the homogeneous ($A=40$), flowing ($A=320$), and vortex states ($A=340$), arranged left to right. Obtained from full nonlinear simulations with $\nu=0$ and $S=1$.}
    \label{fig: schematic_zig_zag}
\end{figure*}

\noindent\textbf{Governing Equations.--} We study a two-dimensional active nematic fluid at low Reynolds number, following the formulation first introduced in \cite{Universal_scaling_AN,Non_Linear_Spontanous_Flow}. Neglecting inertia, the force balance reduces to
\begin{align}
    0 = \partial_\beta \Bigg( & 2 \eta v_{\alpha \beta} -P \delta_{\alpha \beta} +\frac{1}{2}(h_\alpha n_\beta - n_\alpha h_\beta) \notag \\
    & +\frac{\nu}{2}(n_\alpha h_\beta +h_\alpha n_\beta - n_\gamma h_\gamma \delta_{\alpha \beta})- \zeta q_{\alpha \beta} \Bigg),
    \label{eq:force_balance}
\end{align}
subject to the incompressibility condition $\partial_\alpha v_\alpha=0$, where $v_{\alpha \beta}= \frac{1}{2} (\partial_\alpha v_\beta + \partial_\beta v_\alpha)$ is the symmetric part of the strain rate, $q_{\alpha \beta}=n_\alpha n_\beta -\frac{1}{2} \delta_{\alpha \beta}$ is the nematic tensor, and $\sigma^a_{\alpha \beta}= \frac{1}{2} (n_\alpha h_\beta -h_\alpha n_\beta)$ denotes the antisymmetric part of the stress tensor. 

The orientational dynamics of the fluid are governed by the relaxation of a free energy functional, $D n_\alpha/Dt = \frac{1}{\gamma} h_\alpha$, where the free energy takes the form $\mathcal{F}=\frac{K}{2} \int (\partial_\alpha n_\beta)(\partial_\alpha n_\beta) \, d^2 r$ supplemented by a Lagrange multiplier term that constrains the director modulus to $|\hat{n}| = 1$, thereby placing the system deep within the nematic phase. % and inhibiting the appearance of topological defects. This constraint allows us to discard the dynamical equation for the nematic order parameter, thereby simplifying both the numerical and analytical implementation. 
The resulting dynamics of the director field then read

\begin{equation}
    \partial_t n_\alpha + v_\beta \partial_\beta n_\alpha + \omega_{\alpha \beta} n_\beta= \frac{1}{\gamma}h_\alpha -\nu v_{\alpha \beta}n_\beta
\end{equation}
where $\omega_{\alpha \beta}=1/2(\partial_\alpha v_\beta - \partial_\beta v_\alpha) $ is the vorticity tensor, $\gamma$ is the rotational viscosity and $\nu$ represents the flow-alignment parameter, which is positive for disk-like particles and negative for rod-like particles.

Following well-established convention \cite{Universal_scaling_AN,Non_Linear_Spontanous_Flow} we introduce dimensionless variables by rescaling lengths by the system size $L$, time by the active time scale $\tau_a = \eta/|\zeta|$, pressure by the magnitude of the active stress $|\zeta|$, and the molecular field by $K/L^2$. This yields three dimensionless parameters: the activity number $A = L^2/\ell_c^2$, the viscosity ratio $R = \gamma/\eta$, and the sign of the active stress $S = \zeta/|\zeta| = \pm 1$, positive for extensile and negative for contractile stresses. The active length scale $\ell_c = \sqrt{K/(|\zeta| R)}$ is defined by the balance between active and nematic elastic stresses. In terms of these dimensionless variables, we work with the minimal two dimensional active-nematic model for the director angle $\theta$ and stream function $\psi$. To start, we consider the case of zero flow alignment, $\nu = 0$ (see the SM for the full equations), which yields the following equation for the orientation dynamics,
\begin{equation}
    \label{eq:theta}
    \partial_t\theta +(\partial_y \psi)(\partial_x \theta) -( \partial_x \psi)(\partial_y \theta)
    = \tfrac{1}{A}\nabla^2\theta
       - \tfrac{1}{2}\nabla^2 \psi,   
\end{equation}
and for the stream function,
\begin{equation}
    \label{eq:psi}
    \nabla^4\left(\psi+ \frac{R}{2A} \theta\right)
        = 
       - S\Bigl[
           \tfrac{1}{2}(\partial_x^2-\partial_y^2)\sin 2\theta
           - \partial_{xy}^2\cos 2\theta
         \Bigr].
\end{equation}
As we have assumed the system to be inertialess, the dynamics of the stream function are instantaneously slaved to the director angle. Throughout this study, we fix the viscosity ratio to $R=1$.

Linearizing the governing equations (Eqs.~\ref{eq:theta} and~\ref{eq:psi}) about a homogeneous state and introducing a wave-like perturbation reveals the flowing instability (see Fig.~\ref{fig: schematic_zig_zag}), previously established in \cite{Simha2002,R_Voituriez_2005,Giomi_2012}, in which the system spontaneously breaks translational symmetry along one direction via either a splay (for contractile activity $S=-1$) or a bend (for extensile activity $S=+1$) instability.\\ %The steady-state flowing profiles $\theta_s(y)$ and $\psi_s(y)$ are obtained numerically (see End Matter for details). \\
 
 \begin{figure*}[!t]
    \centering
    \includegraphics[width=0.85\linewidth]{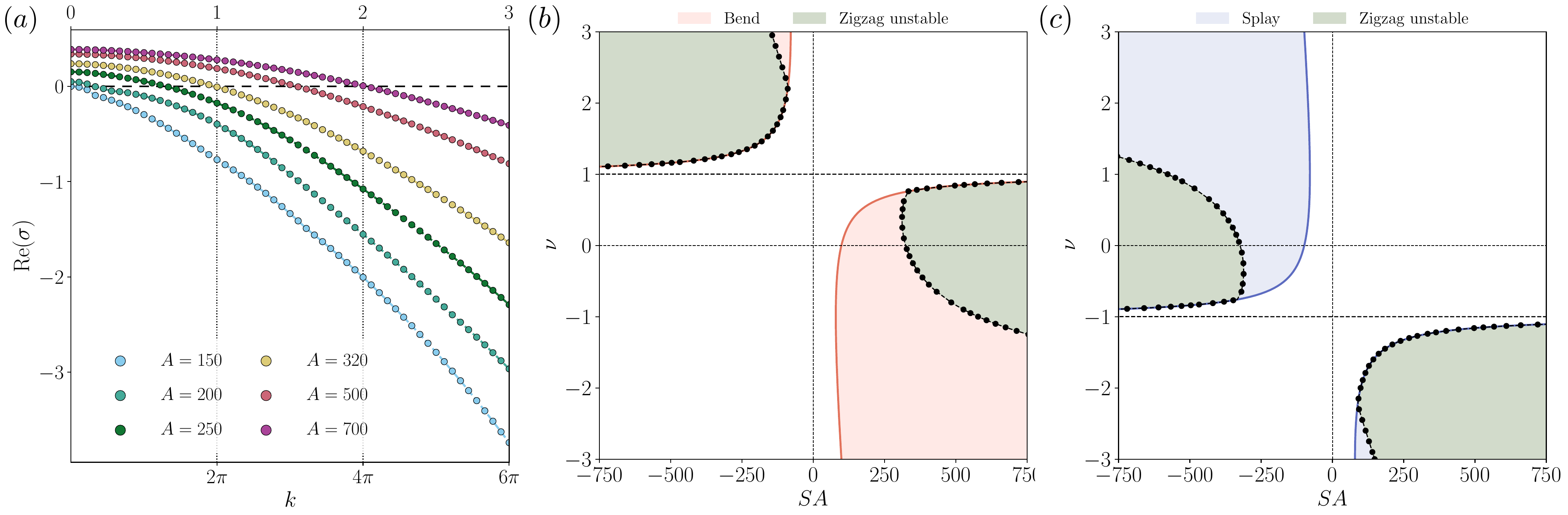}
    \caption{\textbf{ Zigzag instability growth rate and phase diagram}
    \textbf{(a)} Real part of the numerically integrated growth rate, for different activities, of the perpendicular perturbation with wave vector $k$ for $\nu=0$.
    \textbf{(b,c)} Linear stability phase diagram in the $(SA,\nu)$ plane. Solid lines mark the analytical instability thresholds for the bend  and splay modes (Eq~\ref{eq: splay_bend_LSA_lines}), with the corresponding unstable regions shaded accordingly. The scatter points traces the numerically determined zigzag threshold for both contractile ($S=-1$, left) and extensile ($S=+1$, right) activity, with the enclosed zigzag-unstable region shaded in green.
    }
    \label{fig: growth_rate_zig_zag_phase_diagram}
\end{figure*}

\noindent\textbf{Zigzag instability.--} The flowing state consists of a
stripe pattern in the director and stream function, translationally
invariant along the stripe direction (see Fig.~\ref{fig:
schematic_zig_zag}). This one-dimensional pattern remains stable provided
perturbations transverse to the stripes decay. We propose that this
condition fails above a critical activity, and that the resulting
instability, rather than nonlinear dynamics, is what produces the vortex
state. As activity increases, a long-wavelength transverse perturbation can
instead be enhanced. Physically, this perturbation locally bends the
stripes, and because the active stress depends on the local curvature of the
director field, the resulting distortion reinforces itself: regions where
the stripe bends one way generate flows that further displace the pattern in
the same direction, while the alternating curvature along the stripe
produces the characteristic zigzag deformation. Once this transverse mode
grows, the stripes break up into a periodic array of vortices, breaking the
second translational symmetry and marking the transition from the
one-dimensional flowing state to the two-dimensional vortex state (Fig.~\ref{fig:
schematic_zig_zag}).

This transition exhibits a structural analogy to the zigzag instability
first identified in Rayleigh-B\'enard convection, and is likewise captured
within a Floquet stability framework \cite{RevModPhys_Cross,Cross2009}. To
track the onset of the instability quantitatively, we consider the linear
stability of the flowing state $(\theta_s(y), \psi_s(y))$ under
perpendicular perturbations. In this case, the linearization is given by
\begin{align}
    \theta=\theta_s(y)+ \delta \theta(x,y,t) \quad \delta \theta=u(y)e^{\sigma t +i kx},\\ \psi=\psi_s(y)+\delta \psi(x,y,t)\quad \delta \psi=w(y)e^{\sigma t +i kx},\
\end{align}
where $\sigma$ denotes the growth rate of the perturbation, $k$ the wave-vector transverse to the flowing state and $u,w$ are the amplitude envelopes of the perturbation. We relegate the linearized equations with flow alignment to the Supplementary Material. Linearizing the governing equations (Eqs.~\ref{eq:theta} and~\ref{eq:psi}) around the steady flowing states leads to the following equation for the orientation dynamics,
\begin{equation}
  \partial_t\delta\theta
  = \nabla^2 \left(\frac{1}{A}\delta\theta - \frac{1}{2} \delta\psi\right)
    + \partial_y\theta_s\partial_x\delta\psi
    - \partial_y\psi_s\partial_x\delta\theta,
  \label{eq:lin-theta}
\end{equation}
and for the force balance equation we retrieve the following linearized equation,
\begin{equation}
\begin{split}
    \nabla^4 \left( \delta\psi +\frac{R}{2A}\delta\theta\right)\\
    =  -S [(\partial_x^2 - \partial_y^2) &\cos{2 \theta_s}\delta \theta 
    + 2 \partial_x \partial_y( \sin{2 \theta_s} \delta \theta)].
\end{split}
\label{eq:lin-psi}
\end{equation}
Introducing the perturbation ansatz leads to the following set of coupled equations for the orientation dynamics,
\begin{equation}
    \label{eq:lin-u}
    \sigma u
    = (\partial_y^2-k^2) \left(\frac{1}{A} u - \frac{1}{2}w\right)
     + ik [(\partial_y\theta_s) w-(\partial_y\psi_s) u],
\end{equation}
and for the stream orientation coupling,
\begin{equation}
    \label{eq:lin-w}
    \left(\partial_y^2-k^2\right)^2 \left(w+\frac{R}{2A} u\right) 
    = S\mathcal{M}_{k}[u],
\end{equation}
\noindent where $\mathcal{M}_{k}[u]$ is the linearized active stress contribution, which depends on trigonometric functions of the steady flowing state and on the perturbation envelope $u$ (displayed in Eq.~\ref{eq:Mkq} of the End Matter). This yields an eigenvalue problem for the growth rate $\sigma$, with the $y$-dependent perturbation envelopes $u$ and $w$ as the corresponding eigen-functions. To solve this, we expand the envelopes and the stationary flowing solution in Fourier modes, $u(y) = \sum_{m=-N}^{N}u_m e^{i\frac{2\pi m}{\lambda}y}$, $w(y) = \sum_{m=-N}^{N}w_m e^{i\frac{2\pi m}{\lambda}y}$, and $\theta_s(y) = \sum_{n} a_n e^{i\frac{2\pi n}{\lambda}y}$, where $\lambda$ is the period of the stationary solution. Substituting these expansions into the linearized equations, and using Toeplitz coefficient matrices to represent the resulting mode couplings, leads to the following $2(2N+1) \times 2(2N+1)$ generalized eigenvalue problem
\begin{equation}
  \sigma
  \begin{pmatrix}I & 0 \\ 0 & 0\end{pmatrix}
  \begin{pmatrix}\bm{u}\\\bm{w}\end{pmatrix}
  =
  \begin{pmatrix}
    \mathcal{A}_{11} & \mathcal{A}_{12} \\
    \mathcal{B}_{21} & \mathcal{B}_{22}
  \end{pmatrix}  
  \begin{pmatrix}\bm{u}\\\bm{w}\end{pmatrix}.
  \label{eq:gen-eig}
\end{equation}
where the four blocks are obtained from (Eqs.~\ref{eq:lin-u} and~\ref{eq:lin-w}) and read $\mathcal{A}_{11}= \frac{1}{A} \left(\partial_y^2-k^2\right) - ik (\partial_y\psi_s)$, $\mathcal{A}_{12}= ik (\partial_y\theta_s) - \frac{1}{2}\left(\partial_y^2-k^2\right)$, $\mathcal{B}_{21}=\frac{R}{2A}\left(\partial_y^2-k^2\right)^2 -S\mathcal{M}_{k}$ and $\mathcal{B}_{22}=\left(\partial_y^2-k^2\right)^2$. Since $\mathcal{B}_{22}$ is diagonal we can invert and isolate $\bm{w} =-\mathcal{B}_{22}^{-1}\,\mathcal{B}_{21}\bm{u}$. 
Which leads to the reduced ordinary eigenvalue problem,
\begin{equation}
    \label{eq: growth rate matrix no flow aligment}
    \sigma \bm{u} = \mathcal{L}(k)\bm{u} \quad \mathcal{L}(k)=(\mathcal{A}_{11}
      - \mathcal{A}_{12}\,\mathcal{B}_{22}^{-1}\,\mathcal{B}_{21})
\end{equation}
which in the homogeneous limit simplifies to the flowing instability growth rate (see End Matter). The growth rate is then determined by the eigenvalues of $\mathcal{L}(k)$. We perturb the one-stripe stationary state with $\lambda = 1$, which is the state selected by the coarsening dynamics of the flowing state \cite{Universal_scaling_AN}. Evaluating $\mathrm{Re}[\sigma(k)]$ shows that the zigzag instability exhibits a maximal growth rate in the long-wavelength limit (see Fig.~\ref{fig: growth_rate_zig_zag_phase_diagram}a), similarly to the spontaneous-flow instability of active nematics \cite{Simha2002,R_Voituriez_2005}. The first mode compatible with the system size is $k = 2\pi$, which becomes unstable at $A \approx 324$, consistent with the activity value at which the zigzag instability first appears in the full nonlinear simulations (see snapshots in Fig.~\ref{fig: schematic_zig_zag}) . Alternatively, one can treat the wave vector as continuous, so that the marginal-stability condition is evaluated as $k\to0$ rather than at the finite $k$ set by the system size. In this limit, translational symmetry of the flowing state is broken at $A \approx 181$, below this threshold, continuous $x$-translation invariance protects the mode, while above it, the mode becomes the Goldstone mode of the resulting zigzag pattern. The two thresholds therefore differ only through the geometry, since a finite box admits no wavelength longer than its own size. The onset of the zigzag instability depends on the system size as well as on the activity, a dependence that becomes central once the active nematic is confined to a channel, as we show below under the channel instability.\\
 \begin{figure*}[!t]
    \centering
    \includegraphics[width=0.95\linewidth]{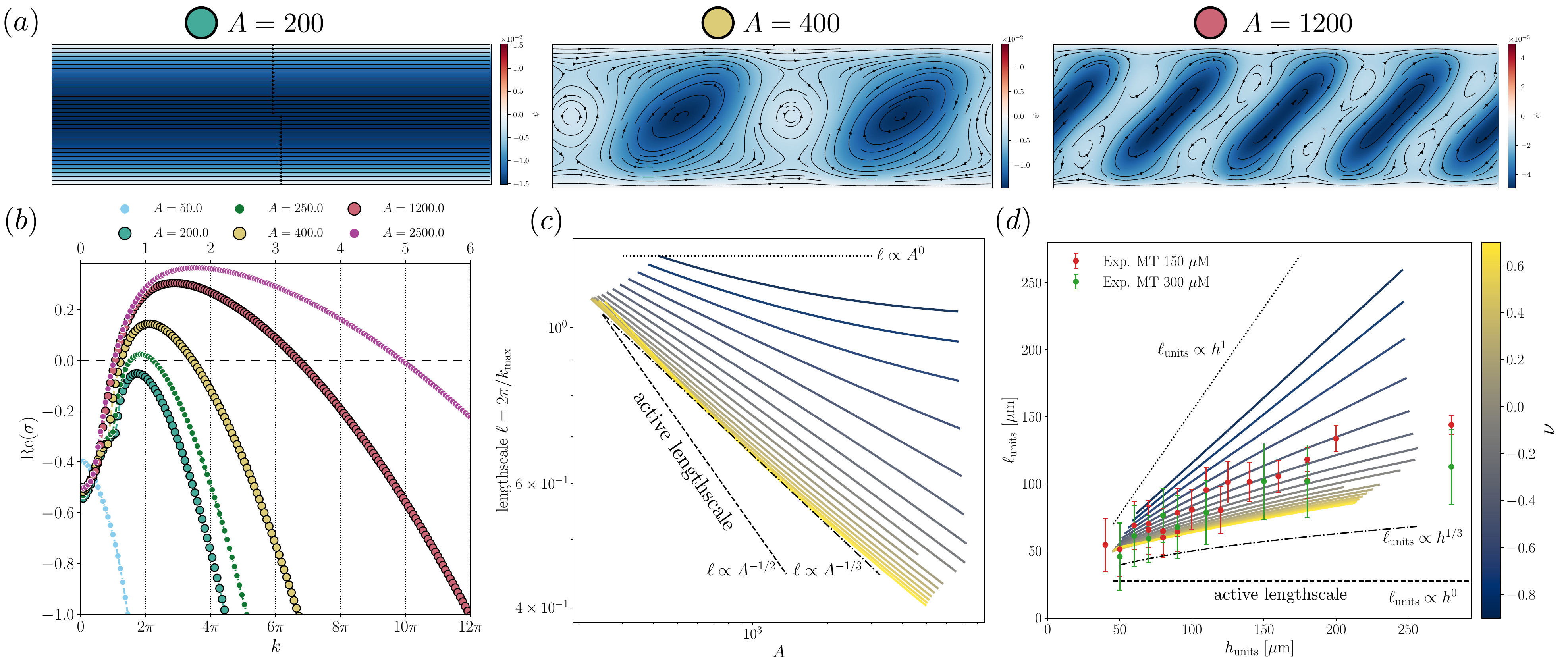}
    \caption{\textbf{The zigzag instability selects an intrinsic length scale along the channel.} \textbf{(a)} Stream function from full nonlinear simulations inside a channel for activities $A = 200, 400, 1200$, with $\nu=0$, $S=-1$ and $L=2$.. From left to right, these show characteristic states of the flowing state, the one pair-vortex state, where the growth rate peaks at $k=2\pi$, and the two-pair vortex state, where the growth rate peaks at $k=4\pi$. \textbf{(b)} Numerically integrated growth rate of a perturbation with wave vector $k$, superimposed on the stationary flowing state for different activity values. \textbf{(c)} Predicted length-scale respect activity for different flow aligning coefficients. \textbf{(d)} Predicted length-scale channel height dependence in $\mu$m. Scatter points indicate the experimental realization of microtubules confined to a channel obtained from Ref. \cite{MT_exp_channel}. Fit uses $\ell_c=3.02 \pm 0.18 \, \mu m$. }
    \label{fig: channel_zig_zag}
\end{figure*}
Repeating this analysis with the flow-alignment parameter $\nu$ included (see SM for details), we determine the critical activity at which the flowing state becomes unstable to perpendicular perturbations, and map the flowing-to-zigzag-unstable transition on the $(SA,\nu)$ phase diagram, adding a new layer to the well-established active nematic phase diagram \cite{Giomi_2012,Edwards_2009,Giomi_2014_Defect_Statistics} (see Fig.~\ref{fig: growth_rate_zig_zag_phase_diagram}b and c). The zigzag-unstable region reveals two qualitative features: (i) For the bend (splay) flowing state, close to the tumbling-to-disk-aligning (tumbling-to-rod-aligning) transition at $\nu = 1$ ($\nu = -1$), the transition to the vortex state becomes direct, and the flowing state itself ceases to exist. (ii) Away from the respective transitions, there exists a regime in which the flowing state persists, and the system must reach a critical activity to trigger the secondary instability. This critical activity increases with $|\nu|$, reflecting the tendency of particles to remain aligned with the flow, which then requires higher energy injection to become unstable under perpendicular perturbations.\\

\noindent\textbf{Channel instability.--} Another hallmark result in the study of active matter is the discovery of the spontaneous flow instability in an active fluid confined to a channel \cite{R_Voituriez_2005,Edwards_2009}. We now examine the zigzag instability in this setting by confining the active nematic within a channel of height $h$ and length $L$. We assume periodic boundary conditions along the $x$-direction of the channel and free slip boundaries at the wall, motivated by experimental observations in microtubule, cells and bacteria \cite{doostmohammadi2018active,MT_exp_channel,wioland2016directed,Spontanous_transition_exp_2018}. 

The confined system analogously exhibits the flowing instability and the longest admissible length scale is selected. %Under the assumed boundary conditions the system can only destabilize the homogeneous state via the splay instability (see the SM for details). 
We perform the same linearization of the equations as in the previous section and perturb the flowing state, recovering the eigenvalue problem stated in Eq.~\ref{eq:gen-eig}, with two qualitative changes arising from the channel geometry: (i) the stationary states $\theta_s$ and $\psi_s$ differ from the unconfined case, and (ii) the perturbation envelopes must instead be expanded in a sine series to satisfy the boundary conditions, $u(y)=\sum_m u_m \sin{\frac{m \pi y}{h}}$ and $w(y)=\sum_m w_m \sin{\frac{m \pi y}{h}}$. Furthermore, we normalize all lengths by the channel height, rather than by the system size as in the unconfined case, as a consequence now the activity number scales with the channel height $A=h^2/\ell_c^2$. The channel length $L$ is then a free parameter (in units of $h$), and the corresponding wave-vector along the channel reads $k = 2\pi n/L$.

The linear stability analysis reveals that the flowing state of a confined system likewise becomes unstable under transverse perturbations, revealing the emergence of vortices, providing mechanistic basis for the numerical and experimental observations of vortex lattice in active nematics \cite{Ceilidh-Dance_2018,Vortex_Lattice} (see simulation snapshots in Fig.~\ref{fig: channel_zig_zag}a). In this case, evaluation of the real part of the growth rate shows that the zigzag instability inside a channel selects an intrinsic length scale (see Fig.~\ref{fig: channel_zig_zag}b). The system becomes unstable at $A \approx 234$, selecting $k=2\pi$ as its characteristic length scale. As activity is further increased, however, a second mode becomes unstable and, at around $A \approx 1200$, comes to dominate over the first, thereby setting the intrinsic length scale and effectively doubling the number of vortices (Fig.~\ref{fig: channel_zig_zag}a). These results provide an analytical account of how increasing activity leads to a growing number of vortices and decrease in the system length scale. Taking the limit of continuous wave vectors, effectively rendering the channel infinite along the $x$-direction, we can select the maximum of the growth rate as the length scale selected by the system at each activity. In Fig.~\ref{fig: channel_zig_zag}c, we display the predicted length scale as a function of the activity coefficient for different flow-alignment parameters, finding a wide range of scalings with activity $l \propto A^{-\alpha}$ and $\alpha$ ranging from $0$ and $1/3$. Furthermore, taking into account the dependence of activity on the channel height, and given that the systems length scale, previously expressed in units of the channel height, now carries dimensions of length, we obtain the following scaling law
\begin{equation}
    \label{eq: scalign_zeta_h}
    l \propto \zeta^{-\alpha} h^{1-2\alpha}.
\end{equation}
When nonlinearities eventually govern the system, the active length scale is recovered ($\alpha = 1/2$) \cite{Universal_scaling_AN,Sumesh_channel_2021}, Eq.~\ref{eq: scalign_zeta_h} predicts that the length scale then becomes independent of the channel height. Notably, none of the predicted scalings are close to the active length scale (Fig.\ref{fig: channel_zig_zag}c). This is indeed in agreement with experimental data from confined bacteria, where the length scale increases with channel height before plateauing (i.e., becoming scale-free) \cite{wioland2016directed}, and with microtubules, where the observed scaling likewise does not correspond to the active-length-scale prediction \cite{MT_exp_channel}. In Fig.~\ref{fig: channel_zig_zag}d, we display the predicted length scale as a function of channel height, together with the experimental data of microtubules confined in a channel adapted from Ref. \cite{MT_exp_channel}, highlighting the appearance of a new length-scale in confined active systems. \\

\noindent\textbf{Discussion.--} Our findings show that the vortex state in active nematics can be understood within linear theory. The spontaneous-flow instability breaks the first translational symmetry, the zigzag instability breaks the second, and the vortex array, usually associated with the nonlinear, turbulent regime, appears while the dynamics is still set by the growth of a single unstable mode. Remarkably, at the tumbling-aligning transitions the two thresholds merge, the flowing state ceases to exist, and the uniformly aligned state becomes unstable directly to the vortex state. Materials in this regime cannot reach the flowing state at any activity, which offers an explanation for why it has proved difficult to observe experimentally, and conversely, whether a given system shows a flowing state constrains its flow-alignment parameter. In a channel, the wavelength of the array depends on both the channel height and the activity, following $l \propto \zeta^{-\alpha} h^{1-2\alpha}$. The exponent $\alpha$ never reaches the value $1/2$ that would reproduce the active length scale, so the selected length scale grows with channel height and does not stay fixed. This is consistent with experimental observations in confined bacterial and microtubule suspensions~\cite{MT_exp_channel,wioland2016directed}. Because the vortex number is quantized and the activity at which each additional vortex pair appears follows from the linear problem, counting vortices while varying the channel height could be used to provide a direct measurement of the
active length scale. Both transitions therefore follow from linear stability analysis, without recourse to the nonlinear dynamics to explain vortex formation in active nematics. \\

\begin{acknowledgments}
We thank Malcolm Hillebrand for helpful discussions, and Jordi Ignés for providing the experimental data on confined microtubule-kinesin motor mixtures. K.T acknowledges funding from the Novo Nordisk Foundation. (grant No.
NNF23OC008501). A. D. acknowledges funding from the Novo Nordisk Foundation (grant No. NNF18SA0035142 and NERD grant No. NNF21OC0068687), Villum Fonden (Grant no. 29476), and the European Union (ERC, PhysCoMeT, 101041418). Views and opinions expressed are however those of the authors only and do not necessarily reflect those of the European Union or the European Research Council. Neither the European Union nor the granting authority can be held responsible for them. The Tycho supercomputer hosted at the SCIENCE HPC center at the University of Copenhagen was used for supporting this work.
\end{acknowledgments}

% \section{Appendix}

\bibliography{apssamp}

@article{Universal_scaling_AN,
	author = {Alert, Ricard and Joanny, Jean-Fran{\c c}ois and Casademunt, Jaume},
	date = {2020/06/01},
	doi = {10.1038/s41567-020-0854-4},
	id = {Alert2020},
	isbn = {1745-2481},
	journal = {Nature Physics},
	number = {6},
	pages = {682--688},
	title = {Universal scaling of active nematic turbulence},
	url = {https://doi.org/10.1038/s41567-020-0854-4},
	volume = {16},
	year = {2020}}

@article{Non_Linear_Spontanous_Flow,
  title = {Nonlinear Spontaneous Flow Instability in Active Nematics},
  author = {Lavi, Ido and Alert, Ricard and Joanny, Jean-Fran\ifmmode \mbox{\c{c}}\else \c{c}\fi{}ois and Casademunt, Jaume},
  journal = {Phys. Rev. Lett.},
  volume = {134},
  issue = {23},
  pages = {238301},
  numpages = {8},
  year = {2025},
  month = {Jun},
  publisher = {American Physical Society},
  doi = {10.1103/PhysRevLett.134.238301},
  url = {https://link.aps.org/doi/10.1103/PhysRevLett.134.238301}
}

@article{Simha2002,
  title = {Hydrodynamic Fluctuations and Instabilities in Ordered Suspensions of Self-Propelled Particles},
  author = {Aditi Simha, R. and Ramaswamy, Sriram},
  journal = {Phys. Rev. Lett.},
  volume = {89},
  issue = {5},
  pages = {058101},
  numpages = {4},
  year = {2002},
  month = {Jul},
  publisher = {American Physical Society},
  doi = {10.1103/PhysRevLett.89.058101},
  url = {https://link.aps.org/doi/10.1103/PhysRevLett.89.058101}
}

@article{R_Voituriez_2005,
doi = {10.1209/epl/i2004-10501-2},
url = {https://doi.org/10.1209/epl/i2004-10501-2},
year = {2005},
month = {apr},
publisher = {},
volume = {70},
number = {3},
pages = {404},
author = {R. Voituriez and J. F. Joanny and J. Prost},
title = {Spontaneous flow transition in active polar gels},
journal = {Europhysics Letters}
}

@article{taming_active_turbulence,
	author = {Guillamat, P. and Ign{\'e}s-Mullol, J. and Sagu{\'e}s, F.},
	date = {2017/09/15},
	doi = {10.1038/s41467-017-00617-1},
	id = {Guillamat2017},
	isbn = {2041-1723},
	journal = {Nature Communications},
	number = {1},
	pages = {564},
	title = {Taming active turbulence with patterned soft interfaces},
	url = {https://doi.org/10.1038/s41467-017-00617-1},
	volume = {8},
	year = {2017}}

@article{MT_Dogic_2012,
	author = {Sanchez, Tim and Chen, Daniel T. N. and DeCamp, Stephen J. and Heymann, Michael and Dogic, Zvonimir},
	date = {2012/11/01},
	doi = {10.1038/nature11591},
	id = {Sanchez2012},
	isbn = {1476-4687},
	journal = {Nature},
	number = {7424},
	pages = {431--434},
	title = {Spontaneous motion in hierarchically assembled active matter},
	url = {https://doi.org/10.1038/nature11591},
	volume = {491},
	year = {2012}}

@article{PRX_turbulence,
  title = {Scaling Regimes of Active Turbulence with External Dissipation},
  author = {Mart\'{\i}nez-Prat, Berta and Alert, Ricard and Meng, Fanlong and Ign\'es-Mullol, Jordi and Joanny, Jean-Fran\ifmmode \mbox{\c{c}}\else \c{c}\fi{}ois and Casademunt, Jaume and Golestanian, Ramin and Sagu\'es, Francesc},
  journal = {Phys. Rev. X},
  volume = {11},
  issue = {3},
  pages = {031065},
  numpages = {16},
  year = {2021},
  month = {Sep},
  publisher = {American Physical Society},
  doi = {10.1103/PhysRevX.11.031065},
  url = {https://link.aps.org/doi/10.1103/PhysRevX.11.031065}
}

@article{MT_toroid,
	author = {Ellis, Perry W. and Pearce, Daniel J. G. and Chang, Ya-Wen and Goldsztein, Guillermo and Giomi, Luca and Fernandez-Nieves, Alberto},
	date = {2018/01/01},
	doi = {10.1038/nphys4276},
	id = {Ellis2018},
	isbn = {1745-2481},
	journal = {Nature Physics},
	number = {1},
	pages = {85--90},
	title = {Curvature-induced defect unbinding and dynamics in active nematic toroids},
	url = {https://doi.org/10.1038/nphys4276},
	volume = {14},
	year = {2018}}

@article{doostmohammadi2018active,
  title={Active nematics},
  author={Doostmohammadi, Amin and Ign{\'e}s-Mullol, Jordi and Yeomans, Julia M and Sagu{\'e}s, Francesc},
  journal={Nature communications},
  volume={9},
  number={1},
  doi={10.1038/s41467-018-05666-8},
  pages={3246},
  year={2018},
  publisher={Nature Publishing Group UK London}
}

@article{Goldstein_2004_Bact,
  title = {Self-Concentration and Large-Scale Coherence in Bacterial Dynamics},
  author = {Dombrowski, Christopher and Cisneros, Luis and Chatkaew, Sunita and Goldstein, Raymond E. and Kessler, John O.},
  journal = {Phys. Rev. Lett.},
  volume = {93},
  issue = {9},
  pages = {098103},
  numpages = {4},
  year = {2004},
  month = {Aug},
  publisher = {American Physical Society},
  doi = {10.1103/PhysRevLett.93.098103},
  url = {https://link.aps.org/doi/10.1103/PhysRevLett.93.098103}
}

@article{bacteria_energy,
  title = {Energy Transport in a Concentrated Suspension of Bacteria},
  author = {Ishikawa, T. and Yoshida, N. and Ueno, H. and Wiedeman, M. and Imai, Y. and Yamaguchi, T.},
  journal = {Phys. Rev. Lett.},
  volume = {107},
  issue = {2},
  pages = {028102},
  numpages = {4},
  year = {2011},
  month = {Jul},
  publisher = {American Physical Society},
  doi = {10.1103/PhysRevLett.107.028102},
  url = {https://link.aps.org/doi/10.1103/PhysRevLett.107.028102}
}

@article{dunkel2013fluid,
  title={Fluid dynamics of bacterial turbulence},
  author={Dunkel, J{\"o}rn and Heidenreich, Sebastian and Drescher, Knut and Wensink, Henricus H and B{\"a}r, Markus and Goldstein, Raymond E},
  journal={Physical review letters},
  volume={110},
  doi = {10.1103/PhysRevLett.110.228102},
  number={22},
  pages={228102},
  year={2013},
  publisher={APS}
}

@article{Spontanous_transition_exp_2018,
	author = {Duclos, G. and Blanch-Mercader, C. and Yashunsky, V. and Salbreux, G. and Joanny, J. -F. and Prost, J. and Silberzan, P.},
	date = {2018/07/01},
	doi = {10.1038/s41567-018-0099-7},
	id = {Duclos2018},
	isbn = {1745-2481},
	journal = {Nature Physics},
	number = {7},
	pages = {728--732},
	title = {Spontaneous shear flow in confined cellular nematics},
	url = {https://doi.org/10.1038/s41567-018-0099-7},
	volume = {14},
	year = {2018}}

@article{Cell_turbulence,
  title = {Turbulent Dynamics of Epithelial Cell Cultures},
  author = {Blanch-Mercader, C. and Yashunsky, V. and Garcia, S. and Duclos, G. and Giomi, L. and Silberzan, P.},
  journal = {Phys. Rev. Lett.},
  volume = {120},
  issue = {20},
  pages = {208101},
  numpages = {5},
  year = {2018},
  month = {May},
  publisher = {American Physical Society},
  doi = {10.1103/PhysRevLett.120.208101},
  url = {https://link.aps.org/doi/10.1103/PhysRevLett.120.208101}
}

@article{extrusion_amin_2018,
	author = {Saw, Thuan Beng and Doostmohammadi, Amin and Nier, Vincent and Kocgozlu, Leyla and Thampi, Sumesh and Toyama, Yusuke and Marcq, Philippe and Lim, Chwee Teck and Yeomans, Julia M. and Ladoux, Benoit},
	date = {2017/04/01},
	doi = {10.1038/nature21718},
	id = {Saw2017},
	isbn = {1476-4687},
	journal = {Nature},
	number = {7649},
	pages = {212--216},
	title = {Topological defects in epithelia govern cell death and extrusion},
	url = {https://doi.org/10.1038/nature21718},
	volume = {544},
	year = {2017}}

@article{flowing_MT,
  title = {Confinement Controls the Bend Instability of Three-Dimensional Active Liquid Crystals},
  author = {Chandrakar, Pooja and Varghese, Minu and Aghvami, S.Ali and Baskaran, Aparna and Dogic, Zvonimir and Duclos, Guillaume},
  journal = {Phys. Rev. Lett.},
  volume = {125},
  issue = {25},
  pages = {257801},
  numpages = {7},
  year = {2020},
  month = {Dec},
  publisher = {American Physical Society},
  doi = {10.1103/PhysRevLett.125.257801},
  url = {https://link.aps.org/doi/10.1103/PhysRevLett.125.257801}
}

@article{Ceilidh-Dance_2018,
	author = {Shendruk, Tyler N. and Doostmohammadi, Amin and Thijssen, Kristian and Yeomans, Julia M.},
	doi = {10.1039/c6sm02310j},
	issn = {1744-683X},
	journal = {Soft Matter},
	month = {06},
	number = {21},
	pages = {3853-3862},
	title = {Dancing disclinations in confined active nematics},
	url = {https://doi.org/10.1039/c6sm02310j},
	volume = {13},
	year = {2017}}

@article{MT_exp_channel,
	author = {Hardo{\"u}in, J{\'e}r{\^o}me and Hughes, Rian and Doostmohammadi, Amin and Laurent, Justine and Lopez-Leon, Teresa and Yeomans, Julia M. and Ign{\'e}s-Mullol, Jordi and Sagu{\'e}s, Francesc},
	date = {2019/10/04},
	doi = {10.1038/s42005-019-0221-x},
	id = {Hardo{\"u}in2019},
	isbn = {2399-3650},
	journal = {Communications Physics},
	number = {1},
	pages = {121},
	title = {Reconfigurable flows and defect landscape of confined active nematics},
	url = {https://doi.org/10.1038/s42005-019-0221-x},
	volume = {2},
	year = {2019}}

@article{Malcolm_Discont_Turbulence,
	author = {Hillebrand, Malcolm and Alert, Ricard},
	date = {2025/12/15},
	doi = {10.1038/s41467-025-67499-6},
	id = {Hillebrand2025},
	isbn = {2041-1723},
	journal = {Nature Communications},
	number = {1},
	pages = {11169},
	title = {Discontinuous transition to active nematic turbulence},
	url = {https://doi.org/10.1038/s41467-025-67499-6},
	volume = {16},
	year = {2025}}

@article{Edwards_2009,
doi = {10.1209/0295-5075/85/18008},
url = {https://doi.org/10.1209/0295-5075/85/18008},
year = {2009},
month = {jan},
publisher = {},
volume = {85},
number = {1},
pages = {18008},
author = {Edwards, S. A. and Yeomans, J. M.},
title = {Spontaneous flow states in active nematics: A unified picture},
journal = {Europhysics Letters}
}

@article{Giomi_2012,
	author = {Giomi, L and Mahadevan, L and Chakraborty, B and Hagan, M F},
	doi = {10.1088/0951-7715/25/8/2245},
	journal = {Nonlinearity},
	month = {jul},
	number = {8},
	pages = {2245},
	publisher = {IOP Publishing},
	title = {Banding, excitability and chaos in active nematic suspensions},
	url = {https://doi.org/10.1088/0951-7715/25/8/2245},
	volume = {25},
	year = {2012}}

@article{alert2022active,
  title={Active turbulence},
  author={Alert, Ricard and Casademunt, Jaume and Joanny, Jean-Fran{\c{c}}ois},
  journal={Annual Review of Condensed Matter Physics},
  volume={13},
  url={https://doi.org/10.1146/annurev-conmatphys-082321-035957},
  number={1},
  pages={143--170},
  year={2022},
  publisher={Annual Reviews}
}

@book{Cross2009,
  author    = {Cross, Michael and Greenside, Henry},
  title     = {Pattern Formation and Dynamics in Nonequilibrium Systems},
  publisher = {Cambridge University Press},
  year      = {2009},
  address   = {Cambridge, UK},
  isbn      = {978-0521770507}
}

@article{RevModPhys_Cross,
  title = {Pattern formation outside of equilibrium},
  author = {Cross, M. C. and Hohenberg, P. C.},
  journal = {Rev. Mod. Phys.},
  volume = {65},
  issue = {3},
  pages = {851--1112},
  numpages = {0},
  year = {1993},
  month = {Jul},
  publisher = {American Physical Society},
  doi = {10.1103/RevModPhys.65.851},
  url = {https://link.aps.org/doi/10.1103/RevModPhys.65.851}
}

@article{wioland2016directed,
  title={Directed collective motion of bacteria under channel confinement},
  author={Wioland, Hugo and Lushi, Enkeleida and Goldstein, Raymond E},
  journal={New Journal of Physics},
  volume={18},
  url = {https://doi.org/10.1088/1367-2630/18/7/075002},
  number={7},
  pages={075002},
  year={2016},
  publisher={IOP Publishing}
}

@article{Vortex_Lattice,
	author = {Doostmohammadi, Amin and Adamer, Michael F. and Thampi, Sumesh P. and Yeomans, Julia M.},
	journal = {Nature Communications},
	number = {1},
	pages = {10557},
	title = {Stabilization of active matter by flow-vortex lattices and defect ordering},
    url={https://doi.org/10.1038/ncomms10557},
	volume = {7},
	year = {2016}}

@article{Giomi_2014_Defect_Statistics,
    author = {Giomi, Luca and Bowick, Mark J and Mishra, Prashant and Sknepnek, Rastko and Cristina Marchetti, M},
    title = {Defect dynamics in active nematics},
    journal = {Philosophical Transactions of the Royal Society A: Mathematical, Physical and Engineering Sciences},
    volume = {372},
    number = {2029},
    pages = {20130365},
    year = {2014},
    month = {11},
    issn = {1364-503X},
    doi = {10.1098/rsta.2013.0365},
    url = {https://doi.org/10.1098/rsta.2013.0365}
}

@article{Sumesh_channel_2021,
    author = {Samui, Abhik and Yeomans, Julia M. and Thampi, Sumesh P.},
    title = {Flow transitions and length scales of a channel-confined active nematic},
    journal = {Soft Matter},
    volume = {17},
    number = {47},
    pages = {10640-10648},
    year = {2021},
    month = {12},
    issn = {1744-683X},
    doi = {10.1039/d1sm01434j},
    url = {https://doi.org/10.1039/d1sm01434j}
}

@article{Jul_Ramaswamy_2016,
	author = {Ramaswamy, Rajesh and J{\"u}licher, Frank},
	date = {2016/02/15},
	doi = {10.1038/srep20838},
	id = {Ramaswamy2016},
	isbn = {2045-2322},
	journal = {Scientific Reports},
	number = {1},
	pages = {20838},
	title = {Activity induces traveling waves, vortices and spatiotemporal chaos in a model actomyosin layer},
	url = {https://doi.org/10.1038/srep20838},
	volume = {6},
	year = {2016}}

@article{living_liquid_crystals,
	author = {Zhou, Shuang and Sokolov, Andrey and Lavrentovich, Oleg D. and Aranson, Igor S.},
	date = {2014/01/28},
	doi = {10.1073/pnas.1321926111},
	journal = {Proceedings of the National Academy of Sciences},
	month = {2026/09/02},
	number = {4},
	pages = {1265--1270},
	publisher = {Proceedings of the National Academy of Sciences},
	title = {Living liquid crystals},
	type = {doi: 10.1073/pnas.1321926111},
	url = {https://doi.org/10.1073/pnas.1321926111},
	volume = {111},
	year = {2014}}

@article{onset_meso_turbulence,
	author = {Doostmohammadi, Amin and Shendruk, Tyler N. and Thijssen, Kristian and Yeomans, Julia M.},
	date = {2017/05/16},
	doi = {10.1038/ncomms15326},
	id = {Doostmohammadi2017},
	isbn = {2041-1723},
	journal = {Nature Communications},
	number = {1},
	pages = {15326},
	title = {Onset of meso-scale turbulence in active nematics},
	url = {https://doi.org/10.1038/ncomms15326},
	volume = {8},
	year = {2017}}

@article{bact_NatPhys,
	author = {Black, Matthew E. and Fei, Chenyi and Alert, Ricard and Wingreen, Ned S. and Shaevitz, Joshua W.},
	date = {2025/09/01},
	doi = {10.1038/s41567-025-02965-y},
	id = {Black2025},
	isbn = {1745-2481},
	journal = {Nature Physics},
	number = {9},
	pages = {1444--1450},
	title = {Capillary interactions drive the self-organization of bacterial colonies},
	url = {https://doi.org/10.1038/s41567-025-02965-y},
	volume = {21},
	year = {2025}}

@article{cancer_firoblast,
	author = {Jacques, C{\'e}cile and Perrin, Louisiane and Ackermann, Joseph and Bell, Samuel and Zajac, Olivier and Lapierre, Ambre and Anger, Lucas and Hallopeau, Cl{\'e}ment and P{\'e}rez-Gonz{\'a}lez, Carlos and Balasubramaniam, Lakshmi and Trepat, Xavier and Ladoux, Beno{\^\i}t and Maitra, Ananyo and Voituriez, Raphael and Matic Vignjevic, Danijela},
	date = {2026/08/01},
	doi = {10.1038/s41563-026-02615-5},
	id = {Jacques2026},
	isbn = {1476-4660},
	journal = {Nature Materials},
	number = {8},
	pages = {1446--1457},
	title = {Fibronectin matrix remodelling modulates the active nematic dynamics of cancer-associated fibroblasts},
	url = {https://doi.org/10.1038/s41563-026-02615-5},
	volume = {25},
	year = {2026}}

@article{AN_Cell_tissues,
author = {Saw, Thuan Beng and Xi, Wang and Ladoux, Benoit and Lim, Chwee Teck},
title = {Biological Tissues as Active Nematic Liquid Crystals},
journal = {Advanced Materials},
volume = {30},
number = {47},
pages = {1802579},
doi = {https://doi.org/10.1002/adma.201802579},
url = {https://advanced.onlinelibrary.wiley.com/doi/abs/10.1002/adma.201802579},
year = {2018}
}

\section*{End Matter}
\appendix

\section{Linearized active stress contribution}

In this section we write the full expression of the linearized contribution of active stress to complement Eq.~\ref{eq:lin-psi},

\begin{align}
   & \mathcal{M}_{k}[u]
   = \cos(2\theta_s)\partial_y^2u \notag\\
    &+ 2\bigl[\partial_y\cos 2\theta_s - ik\sin(2\theta_s)\bigr]\partial_y u 
    \label{eq:Mkq}\\
    &+ \bigl[k^2\cos(2\theta_s)+\partial_y^2\cos 2\theta_s-2ik \partial_y\sin 2\theta_s\bigr]u. \notag 
\end{align}
where $\theta_s$ is the stationary flowing state.

\section{Flowing Instability}

In this section we verify that the linear stability analysis recovers the flowing instability growth rate in the appropriate limit and derive the expression found in \cite{Universal_scaling_AN,Non_Linear_Spontanous_Flow}. Neglecting the transversal perturbation, $k=0$, and setting the stationary states $\theta_s$ and $\psi_s$ to be constants, one recovers, from Eqs.~\ref{eq:lin-theta} and \ref{eq:lin-psi}, the following growth rate,
\begin{equation}
    \label{eq:flowing_growth_rate}
    \sigma = -\frac{1}{A}\left(1+\frac{R}{4}\right)q^2 + \frac{S}{2}\cos{2\theta_s},
\end{equation}
with $q = 2\pi m/\lambda$. This growth rate matches the one obtained for the flowing instability \cite{Universal_scaling_AN}, confirming the consistency of our analysis. Setting the growth rate to zero and including the flow-alignment contributions, we obtain the critical activity required to trigger the flowing instability,
\begin{equation}
    \label{eq: splay_bend_LSA_lines}
    SA = \mp \frac{q^2 \left(4+R(\nu\pm1)^2\right)}{2(1\pm\nu)},
\end{equation}
for the splay and bend instabilities, respectively, which is used in the phase diagrams of Fig.~\ref{fig: growth_rate_zig_zag_phase_diagram}b and c. 

Determining the steady flowing states requires, we set one spatial derivative — in this case, the $x$-derivative — as well as the temporal derivative to zero, yielding, form Eqs.~\ref{eq:theta} and~\ref{eq:psi} the steady flowing states are then given by
\begin{equation}
    \label{eq:stat_flowing}
    \frac{d^2 \theta}{dy^2}=\frac{SA(1+ \nu \cos{2 \theta})(\sin{2 \theta}-\sin{2\theta_0})}{4+R(1+\nu(\nu+2\cos{2 \theta}))}
\end{equation}
 where $\theta_0$ is the orientation of the quiescent state and we have added the flowing alignment contribution. This equation determines the profile $\theta_s(y)$ of the flowing-state solutions for each value of the flow-alignment parameter $\nu$ and activity number $A$ and consequently a flowing stream function $\psi_s(y)$ through Eq. \ref{eq:psi}. 

\section{Numerical Implementation}

We performed full nonlinear simulations using the GitHub repository associated with \cite{Malcolm_Discont_Turbulence}. The director field is evolved with a finite-difference alternating-direction-implicit (ADI) scheme on a periodic $256\times256$ grid, while the force-balance (Stokes) equation for the velocity field is solved spectrally at each time step via FFT. The flowing-profile solutions were obtained using the shooting method employed in \cite{Non_Linear_Spontanous_Flow}. The growth rate evaluation is found solving Eq.~\ref{eq:gen-eig}  projected onto a truncated Fourier basis, with $\sigma$ the eigenvalue of largest real part.

\end{document}